%% file: paper.tex
\documentclass[conference]{IEEEtran}
\IEEEoverridecommandlockouts
\usepackage{cite}
\usepackage{amsmath,amssymb,amsfonts}
\usepackage{algorithmic}
\usepackage{graphicx}
\usepackage{textcomp}
\usepackage{xcolor}
\usepackage{url}
\usepackage{algorithm}

\newif\ifmodify 
\modifytrue 
\ifmodify

\else

\fi

\usepackage{listings}
\usepackage{color}
\definecolor{bgcolor}{RGB}{255, 255, 214}
\definecolor{commentcolor}{RGB}{76,153,0}
\lstdefinelanguage{P4}
{
    morekeywords={parser, if, control, action, apply, table, key, actions, type, header},
    sensitive=true, 
    morecomment=[l]{//}, 
    morecomment=[s]{/*}{*/}, 
    morestring=[b]", 
    morestring=[b]', 
}

\def\BibTeX{{\rm B\kern-.05em{\sc i\kern-.025em b}\kern-.08em
    T\kern-.1667em\lower.7ex\hbox{E}\kern-.125emX}}
\begin{document}


\title{Memory-efficient Sketch Acceleration for Handling Large Network Flows on FPGAs \thanks{This  work  was  funded  by  National  Science  Foundation (NSF) grants CNS-1925464, CNS-1925658, CNS-2130891, CNS-2130907 and CICI-2319962. All opinions and statements in the above publication are of the authors and do not represent NSF positions.}}

\author{\IEEEauthorblockN{Zhaoyang Han\IEEEauthorrefmark{1}, Yicheng Qian\IEEEauthorrefmark{1}
Michael Zink\IEEEauthorrefmark{2}, Miriam Leeser\IEEEauthorrefmark{1}}
\IEEEauthorblockA{Northeastern University \IEEEauthorrefmark{1}, University of Massachusetts Amherst \IEEEauthorrefmark{2}\\
Email: \IEEEauthorrefmark{1}zhhan,qianyich,mel@coe.neu.edu \\
\IEEEauthorrefmark{2}mzink@umass.edu,
}
}

\maketitle

\begin{abstract}
    \input{0-abstract}

\end{abstract}


\section{Introduction}
\input{1-introduction}
\section{Background}
\input{2-background}
\subsection{Related Work}
\input{3-relateWork}
\section{Algorithm and Hardware Implementation}
\input{4-algorithm}


\section{System Implementation}
\input{6-application}

\section{Experiments and Results}
\input{7-experiments}
\section{Discussion and Future Work}
\input{8-discussion}

\section{Conclusion}
\input{9-conclusion}

\bibliographystyle{plain}
\bibliography{P4}

\end{document}

%% file: 0-abstract.tex
Sketch-based algorithms for network traffic monitoring have drawn increasing interest in recent years due to their sub-linear memory efficiency and high accuracy. As the volume of network traffic grows, software-based sketch implementations cannot match the throughput of the incoming network flows. FPGA-based hardware sketch has shown better performance compared to software running on a CPU when handling these packets. Among the various sketch algorithms, Count-min sketch is one of the most popular and efficient. However, due to the limited amount of on-chip memory, the FPGA-based count-Min sketch accelerator suffers from performance drops as network traffic grows. In this work, we propose a hardware-friendly architecture with a variable width memory counter for count-min sketch. Our architecture provides a more compact design to store the sketch data structure effectively, allowing us to support larger hash tables and reduce overestimation errors. The design makes use of a P4-based programmable data plane and the AMD OpenNIC shell. The design is implemented and verified on the Open Cloud Testbed running on AMD Alveo U280s and can keep up with the 100 Gbit link speed.

%% file: 1-introduction.tex
For data processing and analysis, the count-min sketch algorithm, presented by Cormode~\cite{cormode2005improved}, has emerged as a powerful and efficient technique for approximate counting and frequency estimation tasks. It harnesses simple probabilistic hash functions to precisely characterize data sequences while using sub-linear memory space. 


Count-min sketch is widely used in computer networking for tasks such as heavy hitter detection, change detection, and cardinality estimation. The heavy hitter detection problem finds network flows that occupy a significant amount of the network bandwidth. Identifying and taking action against heavy hitters improves network applications like network traffic engineering and can be applied to network intrusion detection~\cite{cormode2005improved,krishnamurthy2003sketch,dobra2002processing}. Field-Programmable Gate Arrays (FPGAs)
are excellent candidates for implementing sketch algorithms. The FPGA architecture is well known for its ability to implement parallelism and pipelining, which is an ideal match for the sketch approach that needs to maintain counters in parallel and process data streams in a pipelined manner.  However, as network traffic proliferates over time and the bandwidth of network interfaces increases, FPGA implementations of the sketch algorithm suffer from performance drops due to limited on-chip memories.    


In this paper, we present a count-min sketch implementation with a variable width counter array designed to provide more efficient memory usage. Hua, et al.\ proposed a variable width counter design: Bucketized Rank Indexed Counters (BRICK)~\cite{hua2008brick} for 64-bit CPUs. We optimize the BRICK architecture for FPGAs. Then we develop a count-min sketch FPGA accelerator using the Protocol-Independent Packet Processing (P4) language combined with the proposed implementation, verify its correct behavior, and measure the performance on an FPGA node directly connected to a 100 Gbps network.

The contributions of this paper are:
\begin{itemize}
    \item We present HBRICK, a hardware friendly memory design implementing variable width counter arrays on FPGA fabric based on BRICK.
    \item We build a data-pipelined count-min sketch architecture with HBRICK. 
    \item We integrate a real time in-network count-min sketch accelerator with the proposed architecture into an FPGA-based NIC (AMD/Xilinx's U280) that proceses 100Gbps packet streams.
\end{itemize}

The resulting design combines P4 and High-Level Synthesis (HLS) to implement a hardware-friendly count-min sketch with a variable width counter and demonstrates the throughput advantage and reduced resource utilization of the design.  The design is tested using network data in the Open Cloud Testbed (OCT)~\cite{zink2021open}.  It demonstrates a design model where the FPGA can directly extract information from the packets and conduct acceleration tasks without the interference of the host. This approach can be extended to other applications where the FPGA processes data directly from the network.  

The rest of this paper is organized as follows. 
Sec.~\ref{sec:background} presents background and related work, focusing on the sketch algorithm with variable width counters and highlighting its benefits. We also discuss the recent works in combining the usage of the network domain-specific language P4 and HLS for developing network functions.
In Sec.~\ref{sec:algo}, we investigate the algorithms of BRICK and our optimized hardware-friendly HBRICK. Sec.~\ref{sec:app} presents the combined use of P4 and HLS to easily implement a system level in-network function design that can be plugged into the FPGA-based SmartNICs. Next, we present our experiments and results. We conclude with a discussion and conclusions. 

%% file: 2-background.tex
\label{sec:background}

\subsection{Sketch with Variable Width Counters}

FPGAs are particularly well suited to accelerate streaming algorithms, where data are streamed into the accelerator and processed in a single pass. As the data size is much larger than the hardware memory resources, the goal is to use sub-linear memory resources compared to the number of data items to estimate a certain variable. 

In this paper, we target a particular streaming data problem: detecting the heavy hitters in network flows. Here, each data item is an individual network packet. We identify a set of packets with the same five-tuple (source and destination IP addresses, source and destination transport protocol number, and transport protocol type) as a packet flow. We characterize a packet flow as a ``heavy hitter'' when the accumulative size of all packets within the flow surpasses a specified threshold~\cite{cormode2005improved}. 

Sketches are algorithms that condense data stream information using sub-linear memory about the data stream size. Typically, sketches comprise multiple counter arrays to approximate the frequency of elements within the stream. In the heavy hitter problem, we use the sketch algorithm to record the packet flow size. While sketches utilize smaller memory, increasing the number of entries in the counter arrays reduces the estimation error of the sketch. With a constrained memory size, reducing the number of bits used for each entry in the counter arrays allows for more entries within each array. This results in better estimation accuracy due to the increased capacity for capturing more elements in the data stream. 

\begin{figure}[ht]
    \center
    \includegraphics[scale=1,width=\linewidth]{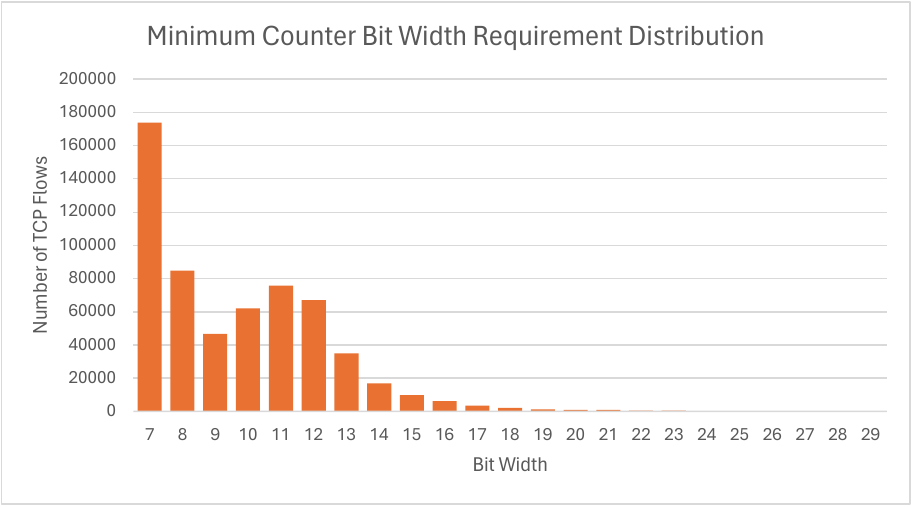}
    \caption{\label{fig:Histo_bits}Minimum Counter Bit Width Requirement Distribution 
of 588K TCP/UDP flows of real network traces. The largest packet flow requires 29-bit to store its total size while most flows only require 14-bit counters.}
\end{figure}

We analyze the sizes of 588K network flows based on real network traces made public by CAIDA~\cite{caida_dataset}. We calculate the minimum bit width required to store these flow sizes.
As shown in Fig.~\ref{fig:Histo_bits}, the real network traces contain mostly small flows. The heavy hitters, which require more memory bits for recording, constitute only a small fraction of the total flows. In other words, only a very small number of counter entries require more bits. One solution to reduce memory usage efficiently is to use variable width counters instead of fixed width. To achieve the variable width counters, extra processing cycles are commonly required to access a certain entry due to its adaptive architecture. Thus, variable width counters provide a trade-off opportunity between computing overhead and memory. 

\subsection{P4-based In-network Computing on FPGA-based SmartNICs}

In-network computing aims to offload compute intensive networking tasks from the CPU to network devices like programmable NICs. In-network computing can reduce the overhead incurred by the CPU host and improve the performance in throughput and other metrics. Different kinds of hardware devices, such as AMD Pensando's DPUs, Nvidia Bluefield SmartNICs, and AMD/Xilnx's FPGAs, have been used for computing in the network. 
There was a lack of device-agnostic high-level abstractions in processing packets among these devices. P4, a domain-specific language originally developed to control packet forwarding in the data plane for devices like routers and switches, provides a language for programming these NIC devices at a higher level~\cite{bosshart_p4_2014}. 

FPGAs have been a good candidate as P4-programmable devices since the beginning of the P4 language. P4-NetFPGA~\cite{ibanez2019p4} created a workflow enabling offloading P4 functions onto FPGA-based NICs while maintaining high throughput. 
To maximize the capabilities of FPGAs in in-network computing, we propose a design that combines P4 and HLS. The intended network task will be segmented into one or more computational kernels, which are constructed using HLS, whereas P4 will control the packet flow through different kernels based on packet headers. We demonstrate our design as a case study of the combination.



%

%% file: 3-relateWork.tex
FPGA-based designs of count-min sketch with its variants have been widely studied in the past decade because of the match between the hardware device features and sketch structures. 
Tong et al.~\cite{tong_sketch_2018,tong_high_2016,tong2013online} implemented FPGA-based count-min sketch in the 2010s. They provide details on count-min
sketch implementations. They analyzed the data conflicts caused by memory access latency and designed a data forwarding unit to mitigate its effect~\cite{tong_sketch_2018}. In contrast,
\cite{chrysos2019data} and \cite{sateesan_speed_2021} describe the design of an approximate counter that can record larger data values with less memory. 
Kiefer et al. ~\cite{kiefer2020scotch} propose a programming model to summarize general sketch designs and their system, Scotch,  generates the hardware designs automatically.
Their techniques are independent of the counter array design, which our work can easily apply. Different groups have also researched other FPGA-based sketch algorithms including Hyperloglog~\cite{kulkarni_hyperloglog_2020,chiosa2021skt, kiefer_optimistic_2023}.

More complicated sketch systems using P4 ~\cite{zhang2021cocosketch, liu2019nitrosketch} implemented on network devices including the Intel Tofino switch~\cite{intel_tofino} have been presented in the past. P4 provides flexibility in designing sketch systems for complex network monitoring tasks. As P4 is device-independent, developers can scale their P4 designs across the network with heterogeneous network devices. In~\cite{sultana2021flightplan}, the authors propose a framework that can disaggregate complex network tasks described by P4 into multiple small functions and assign them to different types of network devices including Intel Tofino switches, AMD's FPGAs and legacy x86 CPUs. Our application also builds on the P4 language, where we use P4 to define the packet-level behaviors for network flows and combine it with HLS to generate a high-performance processing block for count-min sketch that interacts with P4. This allows us to test our design on an FPGA directly connected to the network.  

%% file: 4-algorithm.tex
\label{sec:algo}

\subsection{Count-min Sketch}

In general, a streaming data problem is described as follows: For a given sequence $S_T = \{X_1, X_2, ... X_T\}$, compute a function $F$ of the sequence: $F(S_T)$. 
To solve a data streaming problem while also utilizing sub-linear memory resources, an approximate estimation of the function $F$ is generally used. 

Count-min (CM) sketch is a widely used algorithm to estimate the sum of data elements. The idea is to map a data element to multiple counters using independent hash functions and then increase the value of the corresponding counters by the size of the data element.  To estimate the total size of a certain flow, the algorithm will return the minimum value among all counters for that data element. 

\begin{algorithm}
    \caption{Count-Min Update}\label{alg:cm}
    \begin{algorithmic}
    \REQUIRE {$S_T$}
    \ENSURE {$C$}
    \STATE {$C_d[W] \gets 0$}
    \FOR {$t \gets 1$ to $T$ }
    \STATE{$k_t, c_t \gets X_t$}
    \FOR {$d \gets 1$ to $D$ }
    \STATE {$C_d[h_d(k_t)] = C_d[h_d(k_t)] + c_t$}
    \ENDFOR
    \ENDFOR
    \end{algorithmic}
\end{algorithm}

To illustrate the algorithm, we define $D$ hash functions $h_d \in H$.  Each hash function is associated with a counter with $W$ entries: $C_d[W] \in C$.
The count-min counters are maintained as shown in Algorithm \ref{alg:cm}.
The symbols $k_t$ and $c_t$ are denoted as the key (packet flow ID) and the size of the packet $X_T$. The flow ID can be determined by the five tuples introduced earlier. 

To query the flow information for a given key $k$, the algorithm compares all the corresponding values from all counters and returns the minimum value. It can be described as:
\begin{equation}
F(S_T,k) = min(C_d[h_d(k)]),  \forall d \in D
\end{equation}

The combination of the design parameters $D$ and $W$ defines the estimation error bound of the CM sketch. The theoretical CM error bounds are as follows. For a given $D=log(1/\delta)$, $W=2/\epsilon$ and any threshold $\phi$, with the probability of $1-\delta$, the sketch algorithm will not falsely over-estimate any flow with size lower than $\phi-\epsilon$ of the entire stream. In other words, mathematically, the overestimation error of the count-min sketch will not exceed $1-\delta$. 

\subsection{Challenges in Bucketized Rank Indexed Counters (BRICK)}

\begin{figure}[ht]
    \center
    \includegraphics[scale=1.2,width=\linewidth]{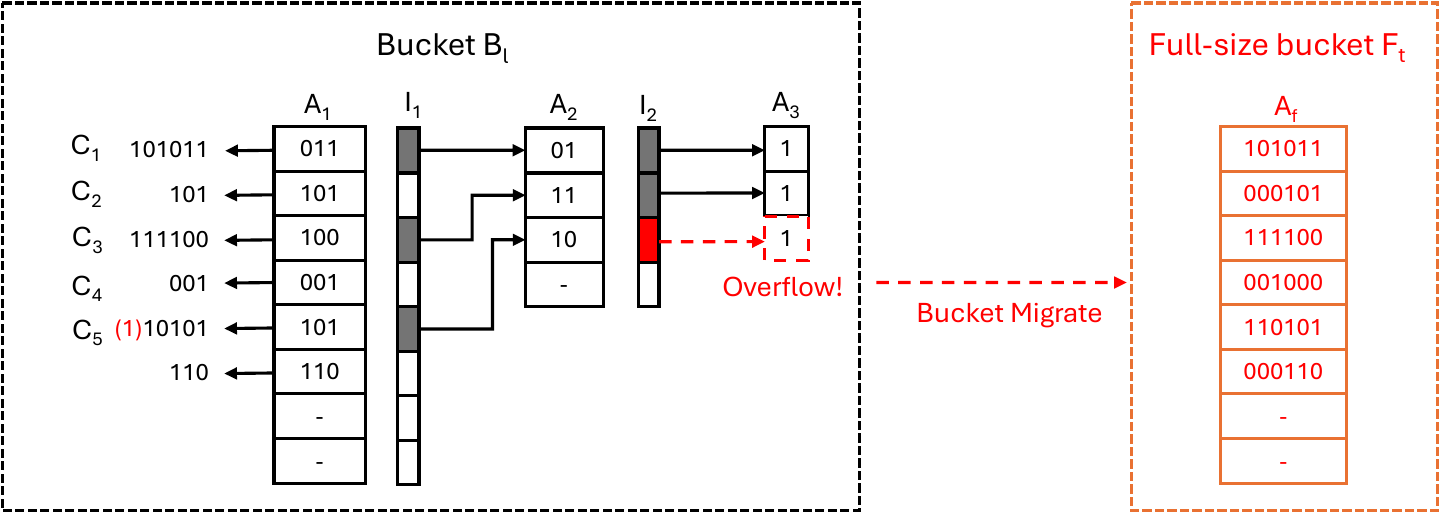}
    \caption{\label{fig:brick_illu}Bucket design of BRICK Architecture and Bucket Migration. The red parts indicate the bucket overflow migration when there is not enough $A_3$ sub-counter bits.}
\end{figure}

BRICK~\cite{hua2008brick}, a variable width counter design, is established based on the idea that not all data entries in the counter will use up all the bits assigned. On the contrary, it assumes that most of the data entries will only require an average number of bits.
The entire BRICK counter system is constructed from $N$ buckets, totaling $W=Nk$ entries. In each counter bucket, the first level contains $k$ sub-counters with average bitwidth, while subsequent levels $A_i$ contain $k_i < k$ sub-counters. We refer to these subsequent level sub-counters as optional levels and $A_1$ as the base level. If we properly select the base level bitwidth as the average bitwidth of the entire data, most of the data will only use up the base level sub-counters. Only a small portion of the data will require optional level sub-counters, i.e. $A_2, A_3, ...$. As a result, we achieve reduced memory utilization through this multi-level memory counter design. Fig.~\ref{fig:brick_illu} depicts a specific three-level counter bucket comprising $k$ counter entries. When an entry with address $C$ is recorded, the correct bucket is determined by $C/k$, and its relative position within the bucket is $C_i = C \bmod k$. The first three bits of the entry are stored in the base sub-counter. If the value exceeds three bits, it automatically overflows to the optional levels.

To connect the data at different levels of the counters, the BRICK approach uses the technique called rank indexing.
Rank indexing is based on 1-bit bitmaps $I$. For an $L$-level design, there will be $L-1$ bitmaps. Each bitmap has the same number of entries as its level. If we denote $I_i$ as the bitmap of i-th level $A_i$, $I_i[p]$ denotes whether the sub-counter entry $A_i[p]$ extends to the next level $A_{i+1}$. If the bit is set to true (shown as a shaded block in the figure), the rank $q = rank(I_{i}[p])$ is entry $p$ 's next-level location. The $rank(I_{i}[p])$ function returns the number of ones in the current bitmap $I_i$ in the range $[0,p]$. 
For example, in Fig.~\ref{fig:brick_illu}, the entry $C_3$ at the first level has its corresponding bit at bitmap $I_1$ set to be true (shaded). It is the second shaded box of the entire bitmap $I_1$, i.e.\ its rank is 2. Therefore the entry $C_3$ has an extension to the next level, $A_2[2]$.

The design divides the counter into small buckets and assumes that data is normally distributed so that data items with large values will fall into different buckets on average. However, real data is often skewed and not normally distributed. Another important scheme of the BRICK design is to handle the insufficient number of sub-counters. As the sub-counters for higher levels are limited, the worst case is the sub-counters at the higher level are not enough to accommodate all the large data items. In other words, more than expected large flows fall into the same bucket. In Fig.~\ref{fig:brick_illu}, the sub-counter $A_3$ overflows as it reaches the capacity of $A_3$.  When this happens, the BRICK design will copy all the data in the bucket to a spare, full width bucket. In~\cite{hua2008brick}, the authors provide concrete mathematical analysis in determining the width and number of entries for each level of the sub counters to reduce the possibility of overflows.

The BRICK design was originally designed for CPU-based sketch counters. When network functions like sketch implementations are offloaded to FPGA-based SmartNICs, it is necessary to accommodate such variable width counter designs onto the FPGA to improve memory efficiency and estimation accuracy. A few challenges need to be addressed for implementing such design on an FPGA to achieve high performance:
\begin{itemize}
\item As multiple cycles are required to read a complete entry from the counters, \textbf{data hazards} are introduced and cause increasing errors in the sketch estimation.  
\item An \textbf{inefficient shift operation} happens during the counter update stage. For example, in Fig.~\ref{fig:brick_illu}, if $C_2$ at level $A_1$ is about to extend to the next level, it should fill the position $A_2[2]$. If $A_2[2]$ is already occupied from earlier counter operations, a shift of $A_2[2]$ and $A_2[3]$ is required. 
\item There is an \textbf{inefficient bucket migration strategy}. The BRICK design requires a few full width buckets and every entry of the bucket has the worst-case width. This is an inefficient use of the memory as all the data in the original bucket have an extra copy in the full-size bucket. In addition, extra time is required for copying the data. 
\end{itemize}

\subsection{Design of Hardware-friendly Bucketized Rank Indexed Counters (HBRICK)}
To address the above issues, we propose the Hardware-friendly Bucketized Rank Indexed Counters (HBRICK).

\begin{figure}[ht]
    \center
    \includegraphics[scale=1.0,width=\linewidth]{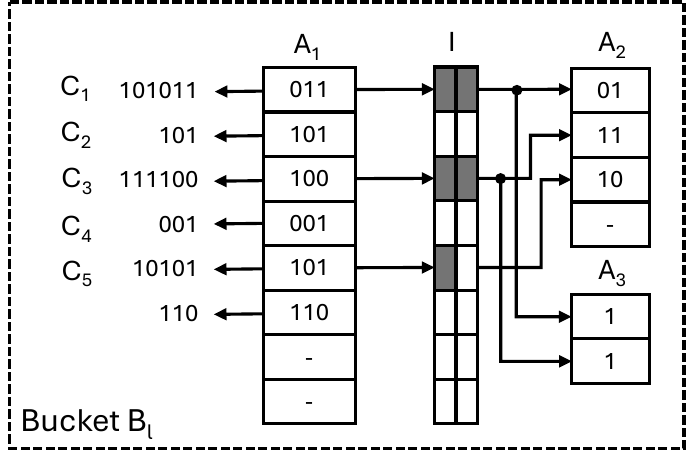}
    \caption{\label{fig:hbrick_illu}HBRICK's optimized bucket design}
\end{figure}

The main issue causing \textbf{data hazards} in the BRICK design is its recursive indexing algorithm, which results in variable cycles to update the counter based on the element's bitwidth. Larger elements with more sub-counters require more cycles. To resolve this, we propose a parallel indexing architecture, as shown in Fig.~\ref{fig:hbrick_illu}. Instead of recursively indexing the sub-counters, we separate the process into an indexing phase and an update phase. We combine the indexing arrays $I_i$ into a single large array $I$. During the indexing phase, We use $I$ to generate the indices for all layers together through the rank operation. With the full-size indexing array $I$, we can index all layers in parallel and fetch the results from sub-counters simultaneously. Our universal index array design simplifies index management. Although the bitmap $I$ requires more memory space than in BRICK, this allows us to control the entire update process at a fixed clock cycle $T_c$, solely based on the number of entries $k$ in each bucket. 
Given this fixed overhead, we create a Data Forwarding Unit (DFU) for each bucket, essentially a FIFO with a depth $T_c$. This unit consolidates multiple accesses to the same entry into a single access over consecutive $T_c$ cycles. Algorithm \ref{alg:read} presents the details of HBRICK.
\begin{algorithm}
    \caption{HBRICK Access}\label{alg:read}
    \begin{algorithmic}
    \REQUIRE {Index $i$}
    \ENSURE {Count $C$}
    \STATE {$C \gets 0$}
    \STATE {Bucket $B \gets i \div N$}
    \STATE{$j \gets i \mod k $}
    \STATE{$S \gets rank(j, I_B)$}
    \FOR {$l \gets 1$ to $L$ }
    \STATE{Note: Iterations execute in parallel}
    \STATE{$C += A_l[S_l]$}
    \ENDFOR
    \end{algorithmic}
\end{algorithm}

To improve the \textbf{inefficient shift operation}, we employ data packing techniques for the optional level sub-counters. We pack all the data in optional levels into single data words respectively, transforming the data shift operation into an atomic bit-wise shift.  While this reduces the total number of entries $k$ in each bucket, it also decreases the processing overhead and helps improve the total throughput. Additionally, since the access for an optional level value is a single word, we can store this word in a local register during consecutive accesses. This approach avoids the data dependency for BRAM reads and writes. However, the drawback of this design is that it is constrained by the maximum bitwidth of the BRAMs, which is 72 bits.

To \textbf{efficiently address bucket overflows}, we propose integrating a BRAM-based fully associative memory block into the design. This addition enables the rapid identification of overflowed entries across the entire counter set. Each entry in the bucket is assigned a dirty bit in an array $V$; this dirty bit indicates whether the entry has been evicted from the bucket. By combining the index array and the dirty bit array, we save additional memory space compared to the original BRICK architecture. According to ~\cite{hua2008brick}, for real network trace with millions of packets, BRICK requires $J=100$ extra full width buckets with $k$ entries each to accommodate overflowed data items. Usually, there are only 1 or 2 overflows that happen per bucket.  As a result, HBRICK requires only $J$ entries in total to migrate the overflows, compared to $k*J$ total entries required in BRICK.  

\begin{figure}[ht]
    \center
    \includegraphics[scale=0.7,width=\linewidth]{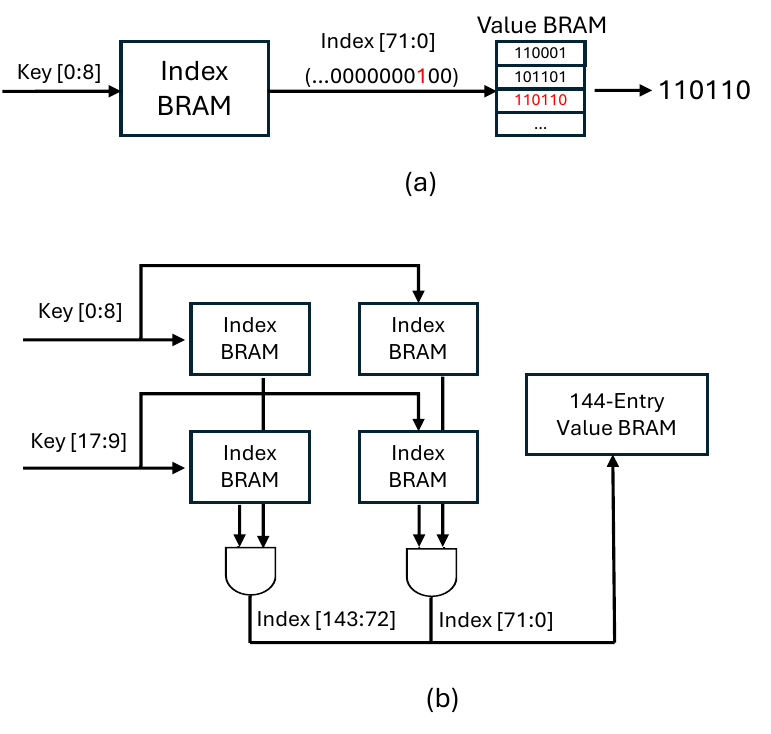}
    \caption{\label{fig:hbrick_assoc}HBRICK Associative Memory Design. (a) demonstrates a single mapping between the 9-bit key and 72 entries. (b) shows an extension of the associative memory with a larger key width and capacity.}
\end{figure}

The design of the associative memory also utilizes the on-chip BRAM. The FPGA dual-port BRAMs are 36Kb memory on AMD Alveo cards and each can be configured as 512 entries with 72-bit wide data. In other words, we can create a simple array with 9-bit addresses to store 72-bit wide data. However, this design will lead to very sparse memory usage. We applied a similar rank index technique used in our bucket design to construct our associative memory since we only need hundreds of entries while the key space is large. For a given key, we use the 72-bit BRAM value to store whether such key is in the associative memory, as shown in Fig.~\ref{fig:hbrick_assoc},  and indicate the location of the value. In HBRICK, we applied the key width based on $W$ and the capacity based on the possibility of overflow~\cite{dhawan2015area}. With the above designs, we demonstrate the counter update algorithm in Algorithm~\ref{alg:write}.

\begin{algorithm}
    \caption{HBRICK Update}\label{alg:write}
    \begin{algorithmic}
    \REQUIRE {Index $i$, New Value C}
    \STATE {Bucket $B \gets i \div N$}
    \STATE{$j \gets i \mod k $}
    \STATE{$S \gets rank(j, I_B)$}
    \IF{WIDTH EXPAND}
        \IF{OVERFLOW}
        \STATE{$V_B[j] \gets 1$}
        \ELSE
        \STATE{Counter Shift: $A[S] >> d_l$}
        \ENDIF        
    \ENDIF
    \STATE{Update Base Sub-counters:}
    \STATE{$A_1[j] = C[:]$}
    \STATE{Update Optional Sub-counters:}
    \FOR {$l \gets 2$ to $L$ }
    \STATE{Note: Iterations execute in parallel}
    \STATE{$A_l[S_l] \&= C[:]$}
    \ENDFOR
    \STATE{Write $A_l$ back to BRAM}
    \end{algorithmic}
\end{algorithm}

\begin{figure}[ht]
    \center
    \includegraphics[scale=1.4,width=\linewidth]{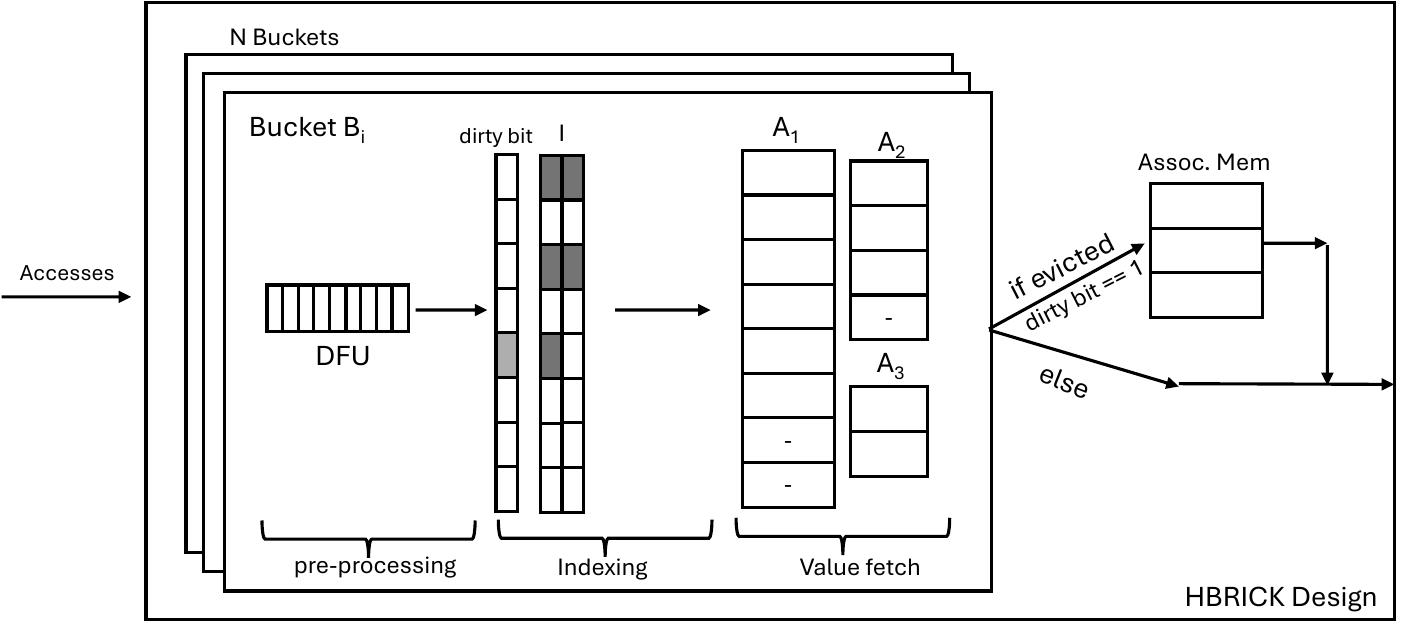}
    \caption{\label{fig:hbrick_sys}Overall HBRICK architecture and data paths. The optimized bucket design includes three stages: pre-processing, indexing, and value fetching.}
\end{figure}

Fig.~\ref{fig:hbrick_sys} shows the overall design of the entire HBRICK counter. The optimized bucket design has three stages. At the pre-processing stage, the access to the same entry will be combined. The indexing stage will calculate the correct indices to access the correct BRAMs among all levels. At last, the values are fetched from the BRAMs and reconstructed into the final value. If the dirty bit is set, access to the associative memory is needed. 

%% file: 6-application.tex
\label{sec:app}

To demonstrate the use of our HBRICK counter in the count-min sketch algorithm and to measure its performance, we develop an in-network count-min sketch implementation with HBRICK and implement it on an FPGA-based NIC. 

We implement the entire system on an FPGA node in the Open Cloud Testbed (OCT)~\cite{zink2021open}, where an AMD Alveo U280 FPGA is used as a SmartNIC attached both to the host CPU and to the 100Gbps network to receive and process incoming packets. The experimental setup is described in Sec.~\ref{sec:exp}. The system is generated with support from the toolchain developed by the OCT operators~\cite{han2023p4-cnert}. The OCT provides a framework that allows network researchers to easily program their P4 applications onto the FPGA through the use of AMD's P4 compiler. In our case, we extended the usage of this framework and combined it with HLS to achieve a much more complicated design. We showcase the usage of two different high-level abstractions under the network processing framework and explore how to connect two different toolchains. 

\begin{figure}[ht]
    \center
    \includegraphics[scale=1,width=\linewidth]{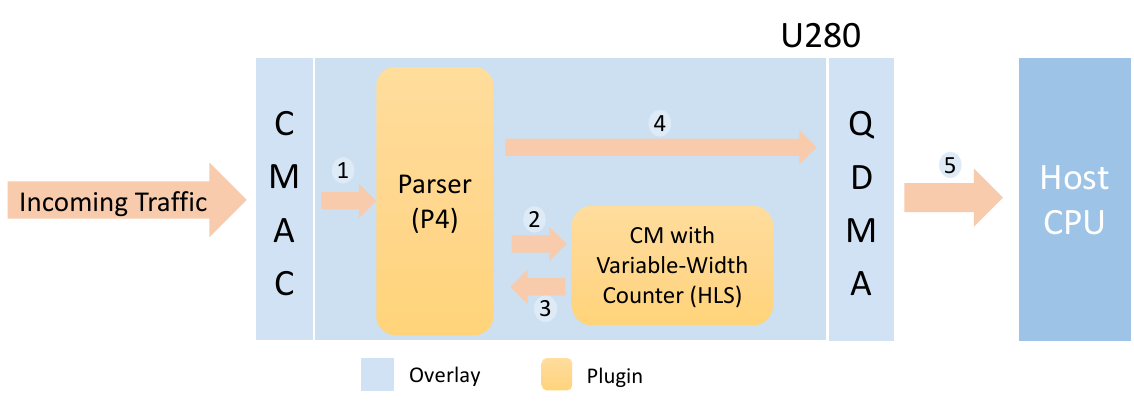}
    \caption{\label{fig:sys} System Architecture for HBRICK-based Count-min Sketch. Data on the numbered path: 1. raw packet; 2. packet key; 3. estimate value; 4\&5 packet forward to the host.}
\end{figure}

This system implementation is broken up into two parts: front end and back end. On the back end, we use the count-min sketch with HBRICK counter generated by HLS as the back end kernel recording and estimating the flows. The front end is a P4 block that controls the packet's path, and extracts header fields as the flow key. It also helps reconstruct the packets and forward the packet to the host.
Fig.~\ref{fig:sys} shows the system architecture of our design. We make use of the open source OpenNIC shell~\cite{opennic} as the fundamental framework, which offers MAC and DMA support for IO. Our application is a plugin block that sits between the CMAC and QDMA blocks of the OpenNIC shell. 


\begin{figure}
\begin{lstlisting}[caption={P4-based front-end}, label=lst:p4_codes,xleftmargin=2.5ex]
...
parser MyParser(packet_in pkt, out headers hdr){
  apply {
    pkt.extract(hdr)};
}
control HHD_Processing(inout headers hdr) {
//  Declare exteranl sketch engine
  *@\textcolor{red}{\textbf{UserExtern<in\_tuples\_t, out\_tuples\_t>(4) sketch;}}@*
...
  apply {
//  Prepare input for the sketch engine
...
    //Call the backend sketch engine
    *@\textcolor{red}{\textbf{sketch.apply(tuple\_in, tuple\_out);}}@*
//  Define post-process behaviours: 
//  Setting the user-defined header field
    if (tuple_out.flow_size > THRESHOLD) 
      hdr.isHitter = 1;
    else 
      hdr.isHitter = 0
    forwardPacket();
  }
} 
...
\end{lstlisting}
\end{figure}
The code snippet for the P4 function is shown in Listing~\ref{lst:p4_codes}. We explore the usage of the P4 extern functions as the connecting point between the HLS-generated blocks and P4 blocks. In particular, we utilize the P4 extern functions as the wrapper of our HLS function and create a standard interface to exchange the necessary information between blocks.
Through the use of the high-level language P4, we have significantly reduced the effort required for designing and implementing network functions. The P4 code requires only about 300 lines, 50\% of which is the definition of headers. Our HBRICK-based count-min sketch also requires about 400 lines of HLS code. The abstraction of the P4 language demonstrates a powerful design model for extracting necessary information directly from the network packets. The FPGA kernel can then bypass the unnecessary data copy in the host CPU. In our case, the extracted information is packet headers. Alternatively,  it can also be information in the packets' payloads. For example, we can easily use this design model to extract images from the network packet and execute machine learning inference on FPGAs directly.

%% file: 7-experiments.tex
\label{sec:exp}
In this section, we describe the testing environment, OCT, and test results of our design. First, we quantify our results in reducing BRAM utilization. Next, we provide a design accuracy profile comparing our HBRICK to the original BRICK. Finally, we present end-to-end throughput results from testing in a real network with actual network packets. 

\subsection{Testing}  For testing, we request an FPGA node from the Open Cloud Testbed~\cite{handagala2022network}. The node is connected to OCT infrastructure through 100G network links for network tests.  We load our implementation onto the FPGA and employ it to receive and identify packets from the network. The packets are real time traces from the Center for Applied Internet Data Analysis (CAIDA) that were collected at the Equinix-Chicago backbone link in January, 2016~\cite{caida_dataset}.
We replay these packets from another OCT node that connects to the same network via 100Gbps port as our FPGA node. The FPGA node processes all the incoming packets from the 100Gb link and then forwards them to the host, where we verify the received packets and measure the performance. The host is running with DPDK \cite{DPDK} as the NIC driver which allows it to keep up with the throughput. 

\subsection{BRAM Utilization and Performance}

We have implemented the design of HBRICK under different configurations to find the optimal choices of the number of layers. We know that increasing the number of counters and layers will result in better estimation accuracy and increased memory utilization. The increased memory utilization will decrease the maximum operating frequency of the module and thus affect the throughput. Also, the increased number of memory layers will lead to large overhead and lower frequency. Therefore, one goal is to determine the optimal trade-off that achieves both high throughput and low estimation error. 

First, we investigate the optimal selection of the number of layers. Table~\ref{table:util} illustrates the impact of the hierarchy of counters on frequencies and estimated processing overheads. As the number of layers increases, memory utilization becomes higher, but this comes at the cost of reduced processing speed and increased overhead. Therefore, we choose the three level HBRICK counters for our performance measurements. 
\input{tables/t_util}

\begin{figure}[ht]
    \center
    \includegraphics[scale=1,width=\linewidth]{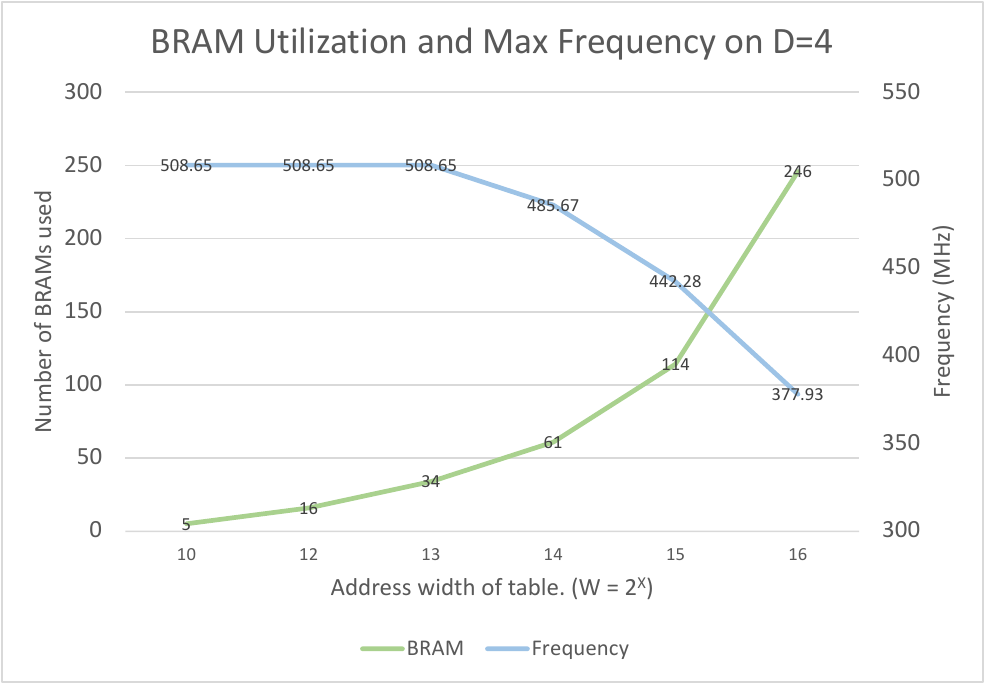}
    \caption{\label{fig:util} BRAM utilization and $F_{max}$ of Our Design with fixed number of hashes $D=4$ while increasing the size (W) of each counter.}
\end{figure}

Fig.~\ref{fig:util} shows a drop in clock frequency as the number of entries in each hash table increases. As the depth of the table increases, the BRAM size also grows. The figure illustrates that when the BRAM exceeds a certain threshold, the operating frequency drops significantly due to the extended data path required to access the larger BRAM. Reshaping the counters array and dividing it into multiple memory banks can slightly increase the frequency, but the clock frequency still experiences a reduction.

\input{tables/t_bram}
Based on the above results, we select $D=4$ and $W=2^{15}$ with three layers $L=3$ of variable counters as our design configuration where the design can run at a maximum 397MHz. It provides a relatively low overestimation error while maintaining a high operating frequency. Using a total of 4 hash functions provides good end-to-end throughput results. Assuming network packets of 64 bytes (the minimum size) as our input stream and that our pipeline can process one network packet in each cycle, our design can achieve a $195Gbps$ throughput with this frequency.

Finally, we implement the other two count-min sketches. One is with the traditional counters and the other is with the original BRICK counters. Table.~\ref{table:bram} shows the BRAM utilization and its max frequency achieved. The table shows that our implementation outperforms the original BRICK in frequency and achieves about $18.9\%$ BRAM utilization improvement.

\input{tables/t_accuracy}
\input{tables/t_compare}

\subsection{Design Accuracy Profiling}

In this section, we use two types of datasets to measure the accuracy and miss rate of our implementation. The first dataset consists of real network traces captured by CAIDA in 2016, containing 3 million packets. The second dataset is synthetically generated to follow the Zipf distribution~\cite{powers1998applications}. For the Zipf distribution, we generate multiple datasets with varying degrees of skewness, determined by the exponent 
$s$. The skewness ranges from 0 to 1.5, with all Zipf datasets containing 3 million packets. As the skewness increases, the total number of flows decreases, resulting in more heavy flows.

We measure the performance of the sketches using the average absolute error, defined as the average error between the real and estimated sizes of all flows. Our design is compared to the traditional count-min sketch (CM), which employs a fixed width counter. Both designs use a similar number of BRAMs.

As shown in Table~\ref{table:acc}, our design demonstrates superior accuracy when the data is skewed. Given the similar total memory usage, overflows are more likely to occur when using the fixed width counters, leading to significant errors. However, as skewness continues to increase, our variable width counter design also shows increased error. When the skewness exceeds 1.25, predominantly heavy flows cause our design to suffer from overflows. Despite this, our design performs well with real world CAIDA datasets and Zipf distribution when the skewness is around 1.


%

\subsection{End-to-End Results}

We implemented the entire design with the selected configurations and tested it on OCT FPGA nodes as described in Sect.~\ref{sec:app}. Our overall design achieves the resource utilization for each block shown in Table.~\ref{table:all_utils}.

\input{tables/t_totutils}
In Table.~\ref{table:compare}, we compare our design with other FPGA implementations. Among all the listed implementations, we measured the real-time throughput through the existing FPGA-based SmartNIC framework. Other implementations provide their throughput based on the generated design frequency. Our measured throughput of this design is about 92Gbps on one Ethernet port of the AMD U280. Our design also shows a comparable design frequency and theoretical maximum throughput with other implementations. 
In \cite{sateesan_speed_2021}, the authors implement an approximate counter design called HSAC, which adopts the Simple Active Counter (SAC), a counter that uses a representation similar to floating-point numbers. As shown in Table.~\ref{table:compare}, our design demonstrates better accuracy compared to this approximation-based counter.
The design presented in~\cite{kiefer_optimistic_2023} shows high performance results using multiple count-min kernels and demonstrating cumulative performance. Their architecture primarily addresses data conflicts between these parallel kernels, using normal counters with universal bitwidth. Our design is orthogonal to their architecture and can further enhance memory efficiency. A detailed discussion will be covered in the next section.

Overall, our design demonstrates comparative performance in accuracy and throughput with reduced memory usage compared to other FPGA-based implementations. 

%% file: tables/t_util.tex
\begin{table}[ht]
\caption{Different HBRICK memory levels for $2^{15}$ entries.}
\label{table:util}
\centering
\begin{tabular}{|l|l|l|l|}
\hline
Levels     & Freq. (MHz)  & Overhead (cycles) & BRAMs        \\ \hline
2          & 412          & 10                          & 132          \\ \hline
\textbf{3} & \textbf{397} & \textbf{14}                 & \textbf{114} \\ \hline
4          & 213          & 18                          & 106          \\ \hline
5          & 74           & 22                          & 99           \\ \hline
\end{tabular}
\end{table}

%% file: tables/t_bram.tex
\begin{table}[ht]
\caption{BRAM utilization comparison between different implementations}
\label{table:bram}
\centering
\begin{tabular}{|l|l|l|l|}
\hline
Implementations   & Freq. (MHz)  & Overhead (cycles)      & BRAMs        \\ \hline
Regular CM        & 453          & 2                      & 153          \\ \hline
Original BRICK    & 156          & 42                     & 103          \\ \hline
\textbf{Ours}     & \textbf{397} & \textbf{14}            & \textbf{114} \\ \hline

\end{tabular}
\end{table}

%% file: tables/t_accuracy.tex
\begin{table}[ht]
\caption{Average absolute error on different datasets}
\label{table:acc}
\centering
\begin{tabular}{|l|l|l|l|l|}
\hline
Datasets & Skewness & CM     & Original BRICK & Ours  \\ \hline
Zipf     & 0        & 1748.4 & 116.7          & \textbf{114.3} \\ \hline
Zipf     & 0.25     & 1712.3 & 119.5          & \textbf{121.6} \\ \hline
Zipf     & 0.5      & 1361.2 & 142.5          & \textbf{142.5} \\ \hline
Zipf     & 0.75     & 1642.5 & 154.3          & \textbf{156.1} \\ \hline
Zipf     & 1        & 472.5  & 125.3          & \textbf{125.3} \\ \hline
Zipf     & 1.25     & 129.6   & 104.4         & \textbf{102.2} \\ \hline
Zipf     & 1.5      & 6.3    & 8.3            & \textbf{7.4} \\ \hline
CAIDA-1  & -        & 432.5  & 105.8          & \textbf{105.3} \\ \hline
\end{tabular}
\end{table}

%% file: tables/t_compare.tex
\begin{table*}[ht]
\centering
\caption{Performance Comparison of Various FPGA-based Designs}
\label{table:compare}
\begin{tabular}{|l|l|l|l|l|l|l|}
\hline
Design                       & Counter Type & Device                           & Theoretical Throughput (Gbps) & Real-time (Gbps) & D, W         & $1-\delta$ \\ \hline
\cite{nguyen2005real}        & exact & Xilinx Virtex 2 XC2V1000         & 61                            & -                & 4, $2^{12}$  & -          \\ \hline
\cite{tong_sketch_2018}      & exact & Xilinx Virtex Ultrascale XCVU440 & 155                           & -                & 5, $2^{16}$  & -          \\ \hline
\cite{tong_sketch_2018}      & exact & Xilinx Virtex Ultrascale XCVU440 & 146                           & -                & 10, $2^{15}$ & 0.999      \\ \hline
\cite{kiefer_optimistic_2023} & exact & Xilinx Alveo U250                & 575 (with three CM kernels)   & -                & -            & 0.99       \\ \hline
\cite{sateesan_speed_2021}    & approx. & Xilinx Virtex UltraScale+        & 196                           & -                & 4, $2^{16}$  & 0.98       \\ \hline
\cite{sateesan_speed_2021}    & approx. & Xilinx Virtex UltraScale+        & 212                           & -                & 5, $2^{15}$  & 0.99       \\ \hline
HBRICK                        & exact & Xilinx  Alveo U280               & 195                           & 92/100           & 4, $2^{15}$  & 0.98       \\ \hline
\end{tabular}
\end{table*}

%% file: tables/t_totutils.tex
\begin{table}[ht]
\centering
\caption{Overall Resource Utilization}
\label{table:all_utils}
\begin{tabular}{|c|c|c|c|c|}
\hline
              & BRAM & LUTs & URAM & DSPs \\ \hline
OpenNIC shell &   16.54\%   &   8\%   &    1\%   &  0    \\ \hline
P4 Front end  &    4.6\%  &  3.4\%    &  0     &    0  \\ \hline
Sketch engine &    30.6\%  &   1\%   &  0     &    0  \\ \hline
\end{tabular}
\end{table}

%% file: 8-discussion.tex
\label{sec:discussion}

\begin{figure}[ht]
    \center
    \includegraphics[scale=1,width=\linewidth]{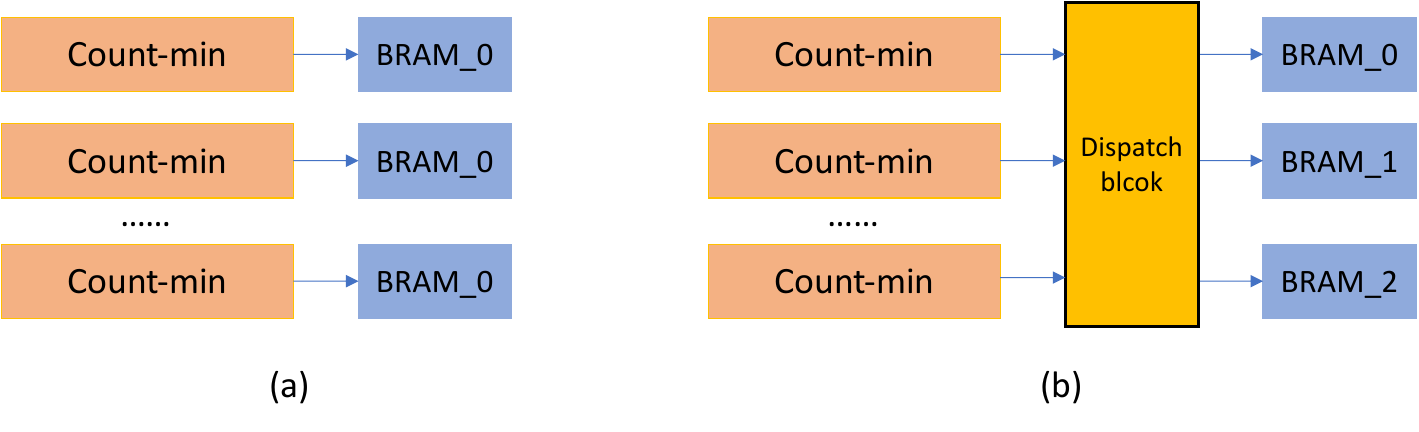}
    \caption{\label{fig:optimisitc} Shared Memory for Parallel Engines. (a) Pessimistic architecture; (b) Optimistic architecture.}
\end{figure}

This paper presents a novel variable width counter architecture for count-min sketch algorithm with reduced memory resource utilization compared to traditional count-min sketch implementations. 
In Sec.~\ref{sec:exp}, we demonstrate the advantage of using our design. Currently, we have implemented a single sketch engine on an Alveo U280 FPGA. An improvement we plan for the future is to increase the number of parallel engines using straightforward duplication.
In~\cite{kiefer_optimistic_2023}, the authors
refer to this as a ``pessimistic'' architecture and they propose an  ``optimistic'' architecture that uses shared memory as shown in Fig.~\ref{fig:optimisitc} with really high-performance as shown in Table.~\ref{table:compare}. Our design is a natural fit for this parallel architecture as we have already divided the counters into numerous small buckets. For future work, we plan to explore our design's performance with parallel engines.

Our design uses P4 controlling packet processing while offloading the sketch task to an external function implemented using HLS.  It demonstrates a design methodology of using the network domain-specific language P4 block to describe network behaviors of packet processing and HLS to achieve high throughput processing units.  This methodology also allows us to orchestrate network flows among multiple HLS-based compute blocks using P4. Another direction for future work is to exploit and design more complicated network traffic monitoring systems with the combination of P4 and HLS.

%% file: 9-conclusion.tex
In this paper we present HBRICK, a novel hardware architecture for count-min sketch with variable width counters, which allow for larger traffic flows to be handled by FPGAs processing data traffic at the edge. The original design for variable width counters ignored data hazards introduced by its recursive update mechanism and inserted extra processing overhead.  In contrast, the HBRICK architecture solves this problem by redesigning the counters and supporting hardware. We implemented our design on FPGAs connected to 100Gb Ethernet network connections. We combine a hardware implementation of the count-min sketch implementation, described in HLS, with a P4 packet processing front end and demonstrate the ability to keep up with the line rate on the network.   

Our implementation and experiments demonstrate one application of network-attached FPGAs with the combination usage of the two high-level abstractions P4 and HLS. Our design constitutes a case study of a potential framework where network-attached FPGA can bypass the host CPU it attaches to and directly extract the necessary information from the network packets through P4 and process them using HLS. 

In the future, we plan to improve the performance of our sketch architecture by exploiting parallelism at several levels. This parallelism will not be limited to a single FPGA, but will be implemented across multiple network-connected FPGAs in the data center to construct a complete networking monitoring system.